# Neutron diffraction evidence for kinetic arrest of the first-order austenite to martensite transition in $Ni_{37}Co_{11}Mn_{42.5}Sn_{9.5}$

V. Siruguri[*], S.D. Kaushik, P.D. Babu,
UGC-DAE Consortium for Scientific Research Mumbai Centre, Bhabha Atomic Research Centre, Mumbai-400085, India

Aniruddha Biswas, S.K. Sarkar, Madangopal Krishnan
Material Science Division, Bhabha Atomic Research Centre, Mumbai-400085, India

P. Chaddah
UGC-DAE Consortium for Scientific Research, University Campus, Khandwa Road, Indore-452001, India

## Abstract

Neutron diffraction measurements, performed in presence of an external magnetic field, have been used to show structural evidence for the kinetic arrest of the first-order phase transition from the high temperature austenite phase to the low temperature martensite phase in the magnetic shape memory alloy $Ni_{37}Co_{11}Mn_{42.5}Sn_{9.5}$ and the formation of a glass-like arrested state (GLAS). The CHUF (cooling and heating under unequal fields) protocol has been used to establish phase coexistence of metastable and equilibrium states of GLAS in the neutron diffraction patterns. We also explore the field-temperature (H,T) phase diagram for this composition which depicts the kinetic arrest line $T_K(H)$. $T_K$ is seen to increase as H increases.

[*]Corresponding author: siruguri@csr.res.in

**Introduction:**

Magnetic first order transformation kinetics of a variety of functional magnetic materials like CMR manganites, magnetic shape memory alloys, intermetallics, multiferroics, etc., have been subject of intense experimental research in the recent times [1-5]. In these materials, typically, the high temperature magnetic phase can be kinetically arrested when cooled under an appropriate external magnetic field, thereby inhibiting the transformation to the low temperature equilibrium phase. This kinetically arrested state has been shown to have coexisting phases of magnetically ordered metastable and equilibrium states and such a material has been termed as a 'magnetic glass' [6,7].

Certain classes of ferromagnetic shape memory alloys (FSMA) like Co-doped NiMnSn, NiMnIn and NiMnAl undergo a kinetic arrest of the first-order austenite to martensite transition [6,8,9]. In these materials, the higher temperature austenite phase has a higher ferromagnetic moment as compared to the lower temperature martensite phase and this manifests itself as a sharp drop in magnetization in a M versus T measurement. By increasing the magnetic field of measurement, it has been observed that the kinetics associated with this first order transformation get hindered and beyond a critical field, they get completely arrested. Hence, there will be regions of field-temperature space where there will be a phase coexistence of metastable (arrested) or glass-like arrested states (GLAS) and equilibrium (transformed) states. If the lower temperature equilibrium (transformed) state has a lower magnetization value than the high temperature austenite phase, it has been noted earlier that by cooling in a certain higher field ($H_c$) and warming in a lower field ($H_w$) would lead to a de-arrest or devitrification of the GLAS [10]. On further warming, the devitrified state would undergo a reverse magnetic and structural transition to the high temperature, high moment austenite phase. This novel protocol of cooling and heating in unequal fields (CHUF) [10] offers an unambiguous method to observe if the glass-like arrested state devitrifies, thereby qualifying it to be called a magnetic glass.

We present here neutron diffraction (ND) studies on the ferromagnetic shape memory alloy $Ni_{37}Co_{11}Mn_{42.5}Sn_{9.5}$ in presence of an external magnetic field over a wide region of the field-temperature (H,T) space and show structural evidence for the kinetic arrest of the first order phase transition from austenite to the martensite phase, subsequent devitrification of the arrested metastable austenite state to the martensite state followed by the reentrant martensite-to-austenite transition on heating from 2K, and finally, phase coexistence of the arrested metastable and equilibrium states by employing the CHUF protocol. Finally, we attempt to explain the (H,T) phase diagram for this composition using the $T_K(H)$ line.

**Experimental details:**

The $Ni_{37}Co_{11}Mn_{42.5}Sn_{9.5}$ buttons were made by vacuum arc melting high purity (99.99%) elements in appropriate proportion. The buttons were solutionized at 1273 K for 24 h in sealed quartz ampoule. High temperature Differential Scanning Calorimetry (DSC) revealed a two-phase melting behavior of this alloy and an extended homogenization treatment was given to ensure structural and chemical homogeneity. Characterization of this alloy has been carried out using Optical and Scanning Electron Microscopy, Electron Probe Microanalysis, XRD and neutron diffraction (ND), Transmission Electron Microscopy (TEM), DSC and DC magnetization. Bulk cylindrical specimens of 5 mm diameter were prepared for ND using electro-discharge machining, as powder samples do not undergo martensitic transformation in this alloy [5]. Nevertheless, one powder specimen was also prepared by crushing a part of the treated button and checked for room temperature structure. The sample for DC magnetization measurements was cut from a different section of the same solutionized button and shaped into 2 mm x 3 mm x 1 mm dimensions. Specimens for TEM were prepared by slicing discs from an electro-discharge machined cylindrical rod of 3 mm diameter, followed by grinding and jet polishing with a Struers Tenupol-5 at 233 K, using a 10 vol % perchloric acid in methanol electrolyte. For ND measurements, the bulk cylindrical specimen was directly attached to the sample stick and inserted into the cryomagnet. The ND patterns were collected using the position-sensitive detector based focusing crystal diffractometer installed by the UGC-DAE CSR Mumbai Centre [11] at the Dhruva reactor, Trombay, at a wavelength of 1.48 Å in the temperature range of 2 K to 300 K under different conditions of applied magnetic field up to 7 Tesla. It was observed that the cylindrical sample consisted of extremely large grains which manifest themselves in almost single crystal-like behavior in the diffraction patterns. Hence, the sample was oriented such that two major reflections of the austenite phase at room temperature, namely, (111) and (200) were clearly visible in the diffraction patterns. The sample was locked in this orientation for all subsequent measurements. DC magnetization was measured using a commercial 9 T PPMS-VSM (make Quantum Design).

**Results and Discussion:**

Alloy characterization:

Fig. 1 displays the high temperature DSC plots of this alloy. It clearly shows the two-phase incongruent melting behavior. Earlier, Watchel et al [12] noted the presence of a peritectic reaction in the ternary $Ni_{0.5}Mn_{0.5-x}Sn_x$ ($0 \leq x \leq 0.5$) pseudo-binary phase diagram. More recently, Yuhasz et al [13] also reported the problem of chemical and structural inhomogeneity that was prevalent in the ternary Ni-Mn-Sn alloys and was caused by this incongruent melting. It was thus, very important to solutionize these alloys for an extended period of time to ensure single-phase and uniform microstructure. Microstructure in as-solutionized condition shows the parent $L2_1$ phase (Fig. 2(a)), as confirmed by the <110> SAD pattern and its simulated key (Figs. 2(b-c)).

DC Magnetization:

Fig. 3 shows the field cooled cooling (FCC) and field cooled warming (FCW) magnetization data at several different fields taken on bulk sample of $Ni_{37}Co_{11}Mn_{42.5}Sn_{9.5}$. The observed thermal hysteresis in the FCC and FCW curves clearly indicates that there is a first order phase transition (FOPT) of high magnetization austenite phase to low magnetization martensite phase in the cooling cycle and the reverse transformation in the warming cycle.

The magnitude of thermal hysteresis gradually increases as the field is raised and has maximum hysteresis for H = 0.5 T, and by 2 T field, the hysteresis disappears as the austenite to martensite transformation is completely hindered by the field and the austenite phase gets kinetically arrested at low temperatures. Such behavior was seen earlier [14] in similar type of compounds. The austenite to martensite transformation start temperature ($M_S$ ~ 181 K) in the cooling cycle and the reverse martensite to austenite transformation finish temperature ($A_F$ ~ 230 K) in the heating cycle, are nearly same for fields up to 0.5 T. For 1.5 T, both these temperatures are shifted towards low temperatures to about 145 K and 214 K, respectively. In other words, the temperature regime over which the hysteresis persists is nearly the same for fields up 0.5 T, and for 1.5 T, it increases significantly although magnitude in terms of moment has come down. Both FCC and FCW curves merge and level off at low temperatures. The magnetization value at 5 K increases systematically with field as it will follow the M-H curve of martensite phase and also the phase fraction of austenite that remains arrested increases with field.

Fig. 4 shows the M versus H isotherms taken at different temperatures. The final temperatures at which MH isotherms were recorded, are reached by cooling the sample directly from 350K in each case. For T > 200 K, the sample is in austenite phase and shows a ferromagnetic MH with high moment value (~130 – 145 emu/gm). At 160 K, the sample is in martensite phase and one observes a ferromagnetic MH isotherm with low moment value (~55 emu/gm). However, with increasing field, one observes a broad field-induced martensite to austenite transformation over a field range of 3 T – 4 T. This trend continues as temperature is lowered to 5 K and only the fields at which this transformation takes place go on increasing. At 5 K, even a field of 9T is not enough to transform the system completely into the austenite phase.

In Fig. 5, the field is raised to 9 T at 350 K and the sample is cooled to 5 K in field. Then, the field is isothermally reduced to zero at 5 K (red curve). This measurement shows that some devitrification of the arrested austenite phase takes place around 5 Tesla (inset of Fig. 5) as indicated by the drop in the magnetization which is still higher than the zero-field cooled (ZFC) magnetization at 5 K. Next, the CHUF protocol was employed to examine the kinetic arrest of the metastable austenite phase and its devitrification to the equilibrium martensite phase and their phase coexistence (Fig. 6). The sample was cooled in different fields ranging from 0.1 T to 8 T down to 5 K and then the field is raised or lowered at 5 K to the value of the measuring field of 0.5 T and the measurements were carried out in the warming cycle. ZFC magnetization curve measured in a field of 0.5 T is also shown for comparison. For cooling fields which are less than 0.5 T, only one transition from a low moment phase to the high magnetization austenite phase is observed which indicates that the low moment phase is the equilibrium martensite phase. Whereas, for coolings fields that are greater than 0.5 T, a sharp drop in the magnetization at low temperatures indicates that a devitrification is taking place from a glass-like arrested phase to an equilibrium martensite phase. On further increasing the temperature, a reverse transformation to the high temperature high moment austenite phase takes place. These observations are discussed along with the neutron diffraction measurements performed using the CHUF protocol in the next section.

Neutron Diffraction:

Neutron diffraction patterns of bulk cylindrical specimens show that the structure could be indexed to an $L2_1$-structure with cell parameter of 5.957 Å (Fig. 7). The bulk sample exhibited strong large grain characteristics behaving almost like a single-crystal. The bulk sample was therefore oriented in such a way that the (111) and (200) reflections of the $L2_1$ austenite phase were strongly visible in the ND pattern. The sample was cooled down to 2 K in this orientation and the peaks associated with the martensite phase appear. Though it is difficult to index the structure with the limited number of reflections, an attempt was made using a standard indexing program and it was observed that the martensite phase is 10M modulated with cell parameters a = 4.338 Å, b = 5.534 Å and c = 21.21 Å and $\beta = 92.55^0$. On warming the sample to 300 K, there is a reverse transformation to the austenite phase in the neighborhood of 230 K. Next, a field of 7 T was employed at 300 K and the sample was cooled again to 2 K and it was observed that the transition to the martensite phase was completely hindered, resulting in a kinetically arrested metastable austenite phase. Employing the CHUF protocol, the sample was cooled in a field of 7 T from 300 K several times but warmed in different fields each time. The warming fields used were 0.5 T, 1.0 T, 1.5 T and 2 T. When the field was reduced from 7 T to 0.5 T for the first warming cycle, it was observed that the arrested austenite phase starts to devitrify at 2 K in a field of 0.5 T itself (Fig. 8(a)). In the next CHUF cycle, when the field was reduced from 7 T to 1 T at 2 K, the sample remains in a kinetically arrested state at 2 K (Fig. 8(b)), as evident from the absence of any peaks attributable to martensite phase. Devitrification sets in at around 10 K and progresses with increasing temperature up to 230 K, beyond which the system re-enters the austenite phase. With warming fields of 1.0 T and above (Figs. 8(b-d)), it is seen that the temperature at which the devitrification sets in, increases with increase in the warming field. However, it is clear from Fig. 3, that cooling and warming in fields greater than 2 T would cause the transition to be completely arrested and no devitrification would be observed at any temperature below the standard martensite to austenite transition temperature. It is important to note here that the quantity ($H_c$-$H_w$) which is difference between the cooling field ($H_c$) and the warming field ($H_w$) is always positive and in the present case, it becomes a precondition to (i) observe the de-arrest of the metastable austenite phase which is similar to the phenomenon of devitrification of a conventional glass upon heating and (ii) the transformation of the devitrified equilibrium state to the high temperature austenite phase which is analogous to melting of devitrified glass. In Fig. 9, the temperature $T_K$ at which devitrification, marked by the appearance of Bragg reflections attributable to the equilibrium martensite phase, sets in, is plotted against the value of the warming field $H_w$. It is observed that $T_K$ increases with $H_w$. The set of temperature points $T_K$ collectively form the kinetic arrest line which, when traversed across in the heating cycle, would result in the de-arrest of the metastable austenite phase to the equilibrium martensite phase. The line also indicates that at lower warming fields, one would have a higher phase fraction of the martensite phase when compared to higher warming fields. In order to substantiate this, the integrated intensity of $(103)_M$ reflection at 140 K, which is the temperature at which the martensite phase fraction is expected to be at its peak value, is plotted as a function of $H_w$ which is the field in which the ND patterns are recorded in each warming cycle (Fig. 10). Expectedly, the martensite phase fraction decreases at higher warming fields.

In conclusion, we present here the first direct neutron diffraction evidence, using the CHUF protocol, for:

(i)  Observation of kinetic arrest of the austenite to martensite first order phase transformation in a shape memory alloy

(ii) Devitrification of the arrested metastable austenite phase as the sample is warmed, followed by the reverse transformation from the martensite to austenite phase at a higher temperature

(iii) Increase in the devitrification temperature $T_K$ as the warming field increases.

**References:**

1. Kranti Kumar, A.K. Pramanik, A. Banerjee, and P. Chaddah, Phys. Rev. B73, 184435 (2006).
2. M.A. Manekar, S. Chaudhary, M.K. Chattopadhyay, K.J. Singh, S.B. Roy, and P. Chaddah, Phys. Rev. B64, 104416 (2001).
3. K. Sengupta and E.V. Sampathkumaran, Phys. Rev. B73, 20406 (2006).
4. A. Banerjee, P. Chaddah, S. Dash, Kranti Kumar, Archana Lakhani, X. Chen and R.V. Ramanujan, Phys. Rev. B84, 214420 (2011).
5. R.Y. Umetsu, K. Ito, W. Ito, K. Koyama, T. Kanomata, K. Ishida and R. Kainuma, J. Alloys and Compounds 509, 1389 (2011).
6. Archana Lakhani, A. Banerjee, P. Chaddah, X. Chen and R.V. Ramanujan, J. Phys. Condens. Matter 24, 386004 (2012) and references therein.
7. P. Chaddah and A. Banerjee, arXiv:1004.3116 and references therein.
8. X. Xu, W. Ito, M. Tokunaga, R.Y. Umetsu, R. Kainuma, K. Ishida, Mater. Trans. 51, 1357 (2010).
9. J.L. Sanchez Llamazares, B. Hernando, J.J. Sunol, C. Garcia and C.A. Ross, J. Appl. Phys. 107, 09A956 (2010).
10. P. Chaddah and A. Banerjee, arXiv:1201.0575 and references therein.
11. A.V. Pimpale, B.A. Dasannacharya, V. Siruguri, P.D. Babu, and P.S. Goyal, Nucl. Instrum. Methods A481, 615 (2002)
12. E. Watchel, F. Henninger and B. Predel, J. Magn. Magn. Mater. 38, 305 (1983)

13. W.M. Yuhasz, D.L. Schlagel, Q. Xing, R.W. McCallum and T.A. Lograsso, J. Alloys Comp. 492, 681 (2010)
14. A. Banerjee, S. Dash, Archana Lakhani, P. Chaddah, X. Chen and R.V. Ramanujan, Solid State Communications 151, 971 (2011).

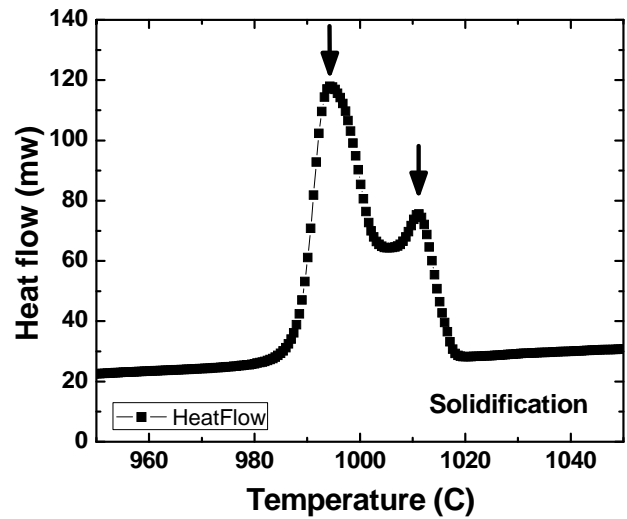

**Fig. 1**. High temperature DSC plots of $Ni_{37}Co_{11}Mn_{42.5}Sn_{9.5}$ showing the two-phase incongruent melting behavior.

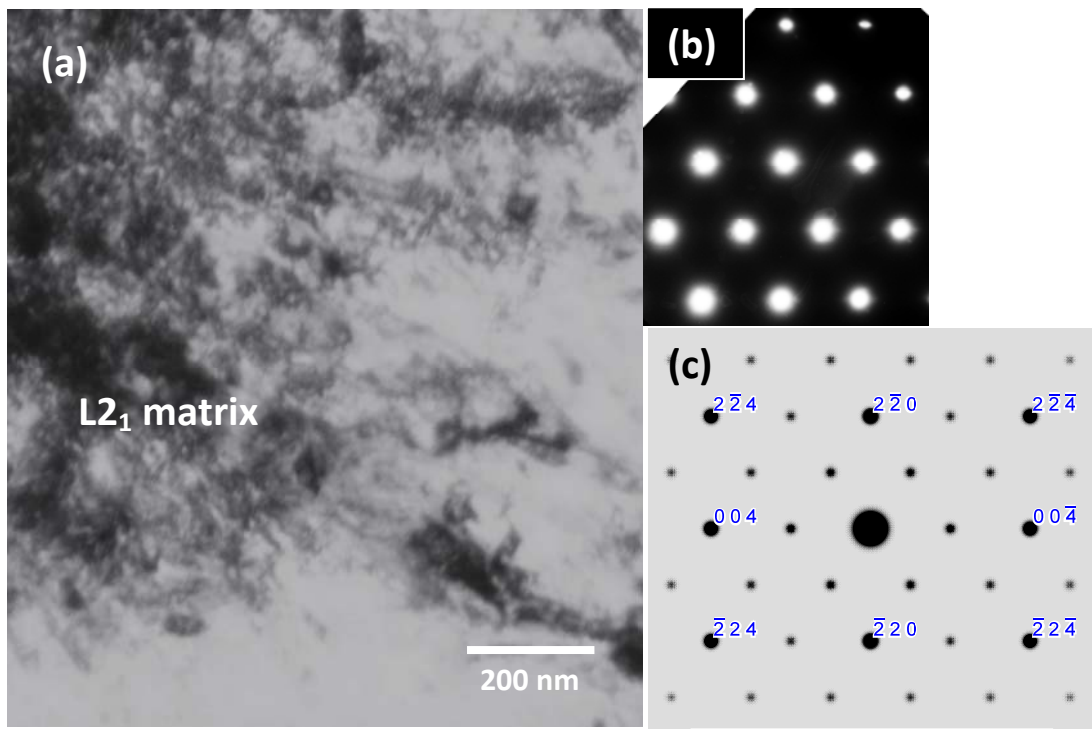

**Fig. 2(a)**. Microstructure of the parent $L2_1$ phase in as-solutionized condition, **(b)** SAD pattern of the <110> zone axis, and **(c)** its simulated key.

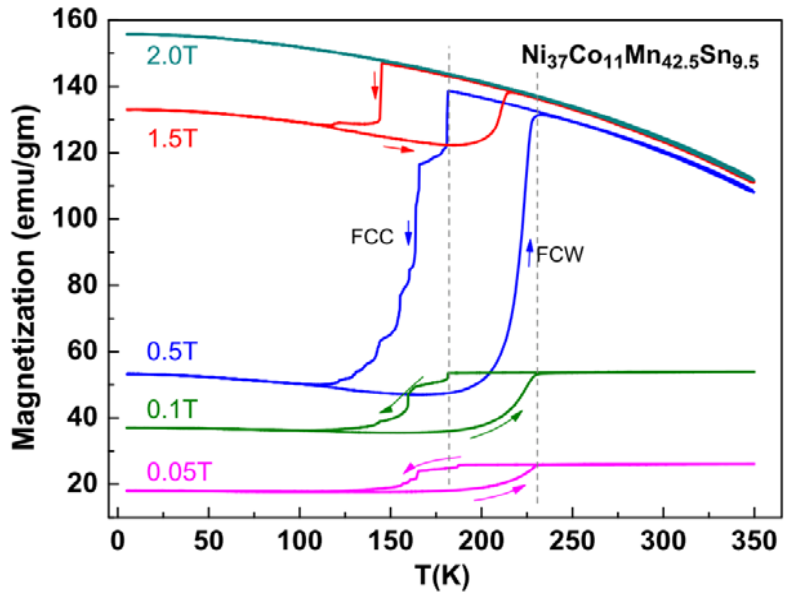

**Fig. 3**. M versus T plots showing the field-cooled cooling (FCC) and field-cooled warming (FCW) curves in applied fields of 0.05, 0.1, 0.5, 1.5 and 2 Tesla.

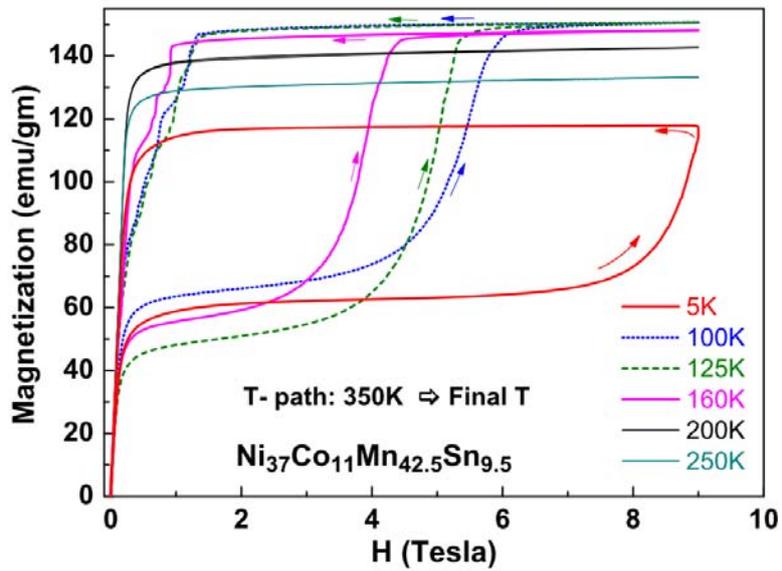

**Fig. 4.** M versus H isotherms at various temperatures after cooling the sample from 350 K.

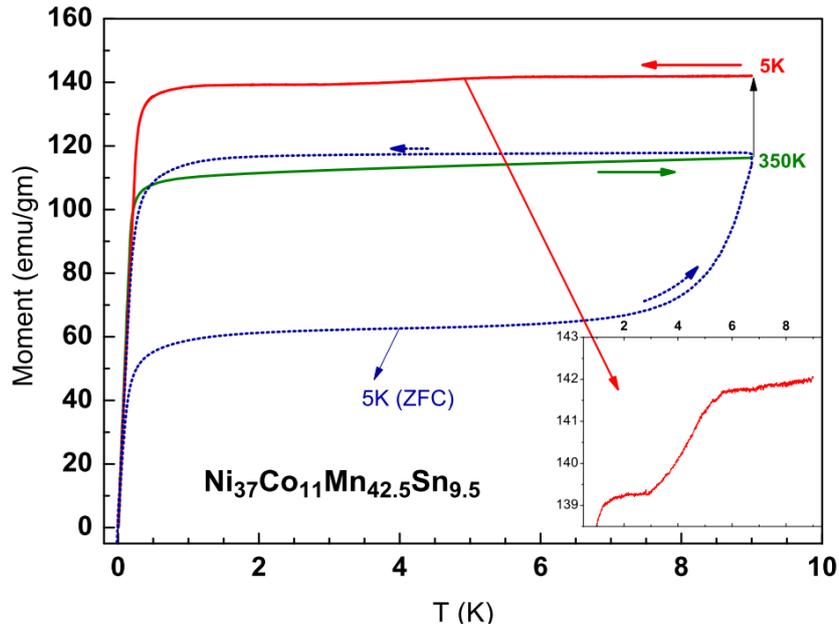

**Fig. 5.** Field-cooled MH isotherms where field is raised to 9 T at 350 K (green curve) and sample cooled to 5 K in field after which the field is reduced to zero (red curve) during measurement. ZFC-MH (blue curve) at 5 K is also shown for comparison. Inset shows the magnified part of FC-MH at 5 K, which shows the devitrification of the arrested high temperature phase.

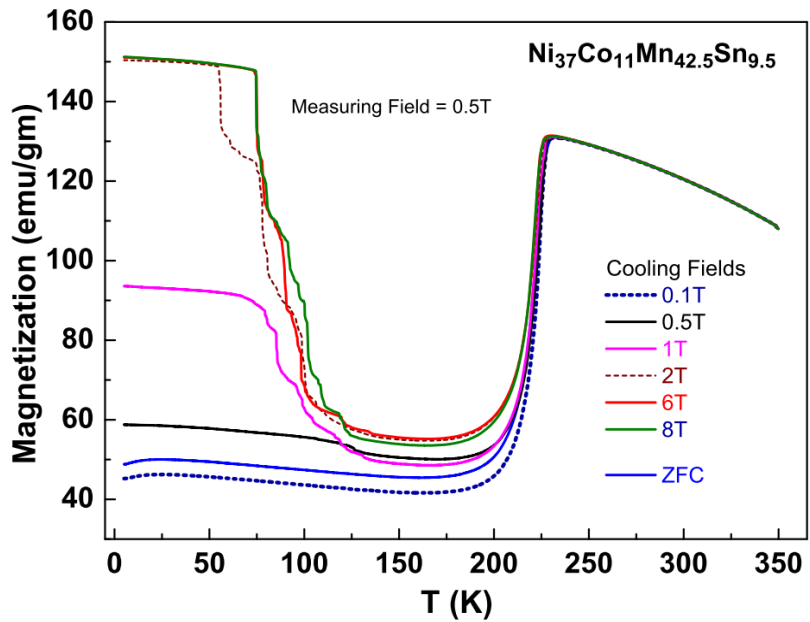

**Fig. 6.** Magnetization as a function of temperature using the CHUF protocol. The sample is cooled in various fields and measured in 0.5 T in the warming cycle.

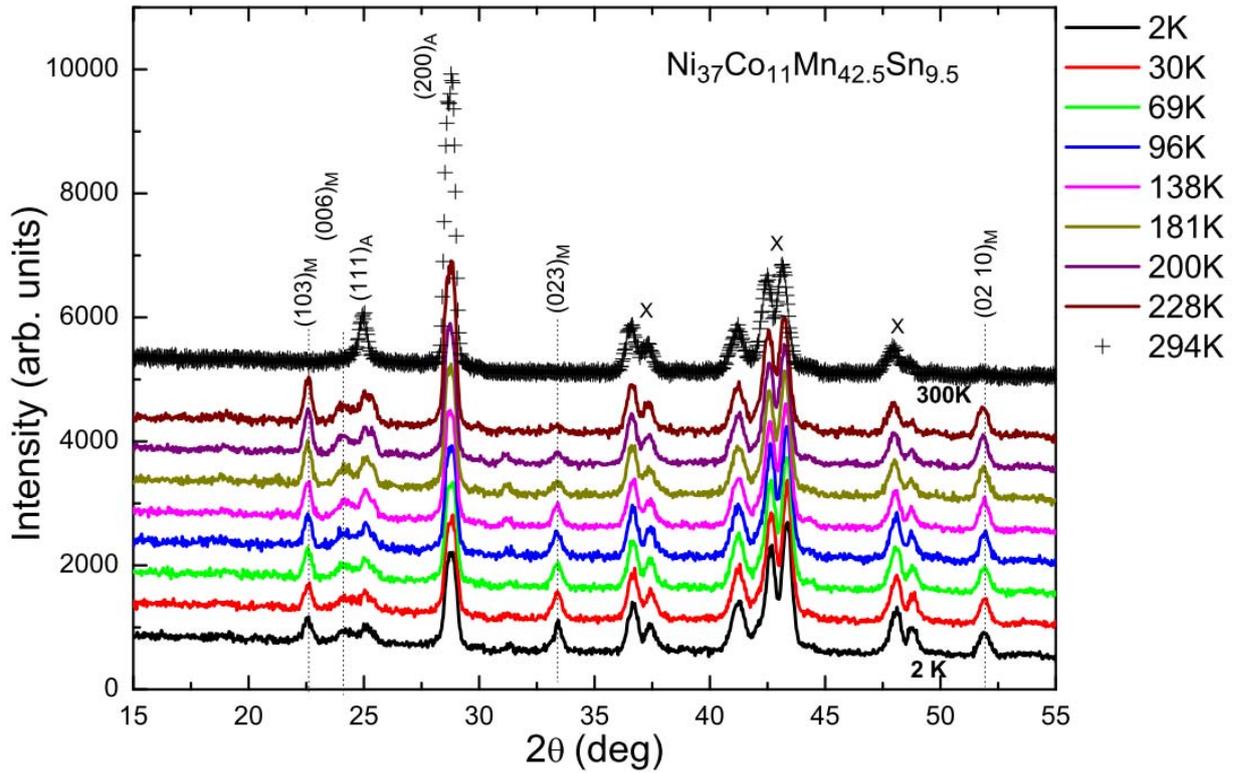

**Fig. 7.** Neutron diffraction patterns as a function of temperature. Indices with subscript 'A' belong to the austenite phase while those with subscript 'M' belong to the 10M martensite phase. Peaks in the regions marked as 'X' are contributions from the magnet shroud. Patterns have been offset for clarity.

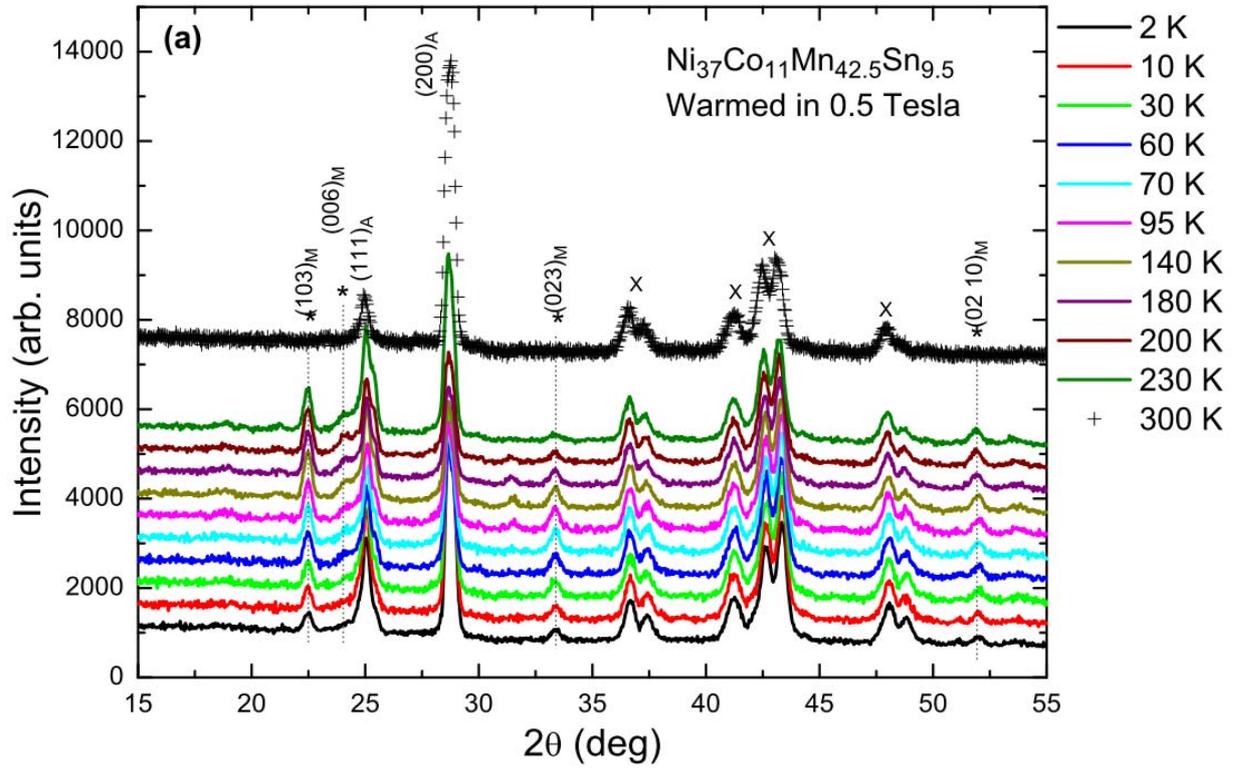

**Fig. 8(a).** Neutron diffraction patterns obtained using the CHUF protocol. The sample was cooled from 300 K to 2 K in a field of 7 Tesla which is isothermally reduced to 0.5 T. Data were taken in the warming cycle. Subscripts A, M and X have the same meaning as in Fig. 7.

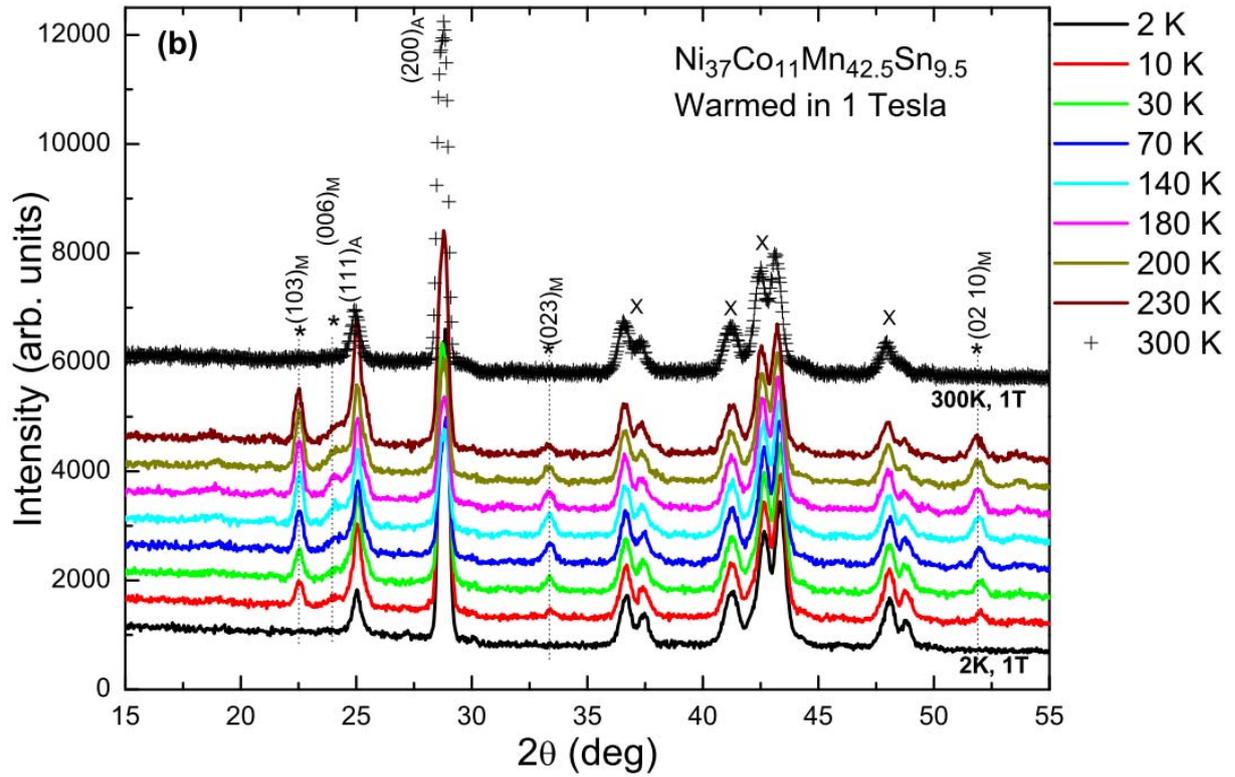

**Fig. 8(b).** Neutron diffraction patterns obtained using the CHUF protocol. The sample was cooled from 300 K to 2 K in a field of 7 Tesla which is isothermally reduced to 1 T. Data were taken in the warming cycle. Subscripts A, M and X have the same meaning as in Fig. 7.

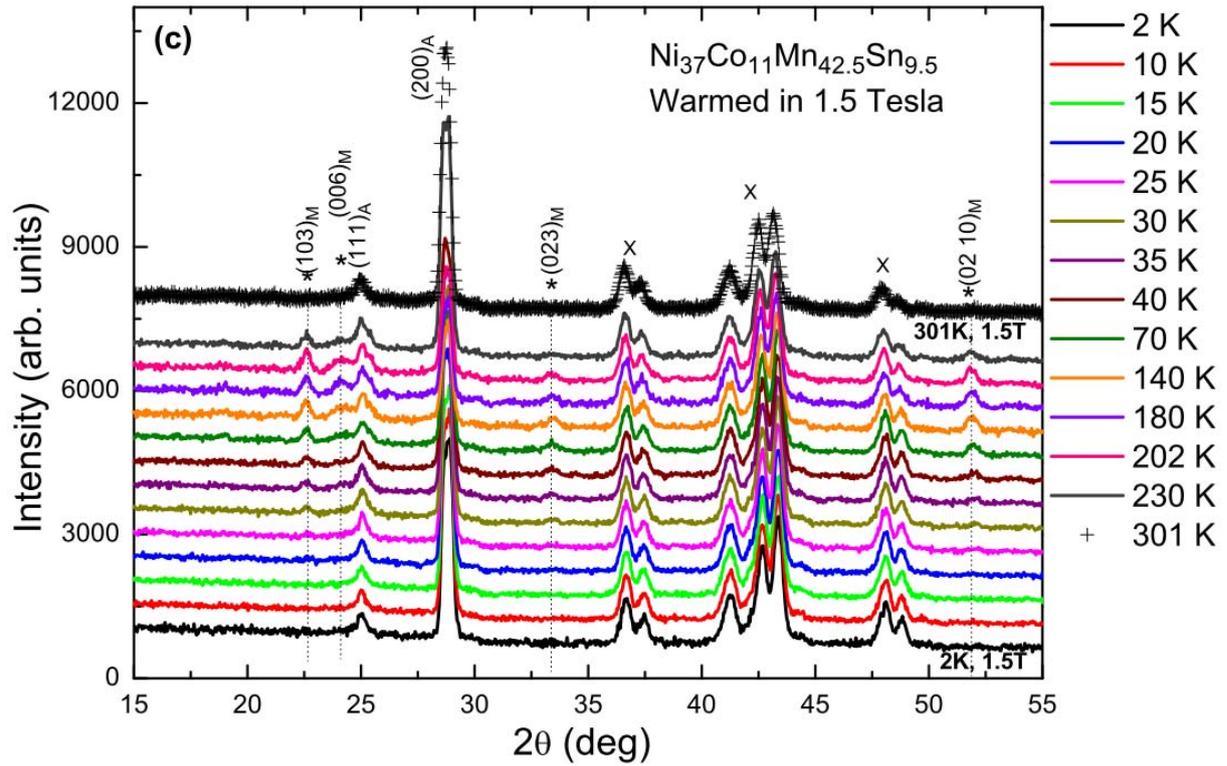

**Fig. 8(c).** Neutron diffraction patterns obtained using the CHUF protocol. The sample was cooled from 300 K to 2 K in a field of 7 Tesla which is isothermally reduced to 1.5 T. Data were taken in the warming cycle. Subscripts A, M and X have the same meaning as in Fig. 7.

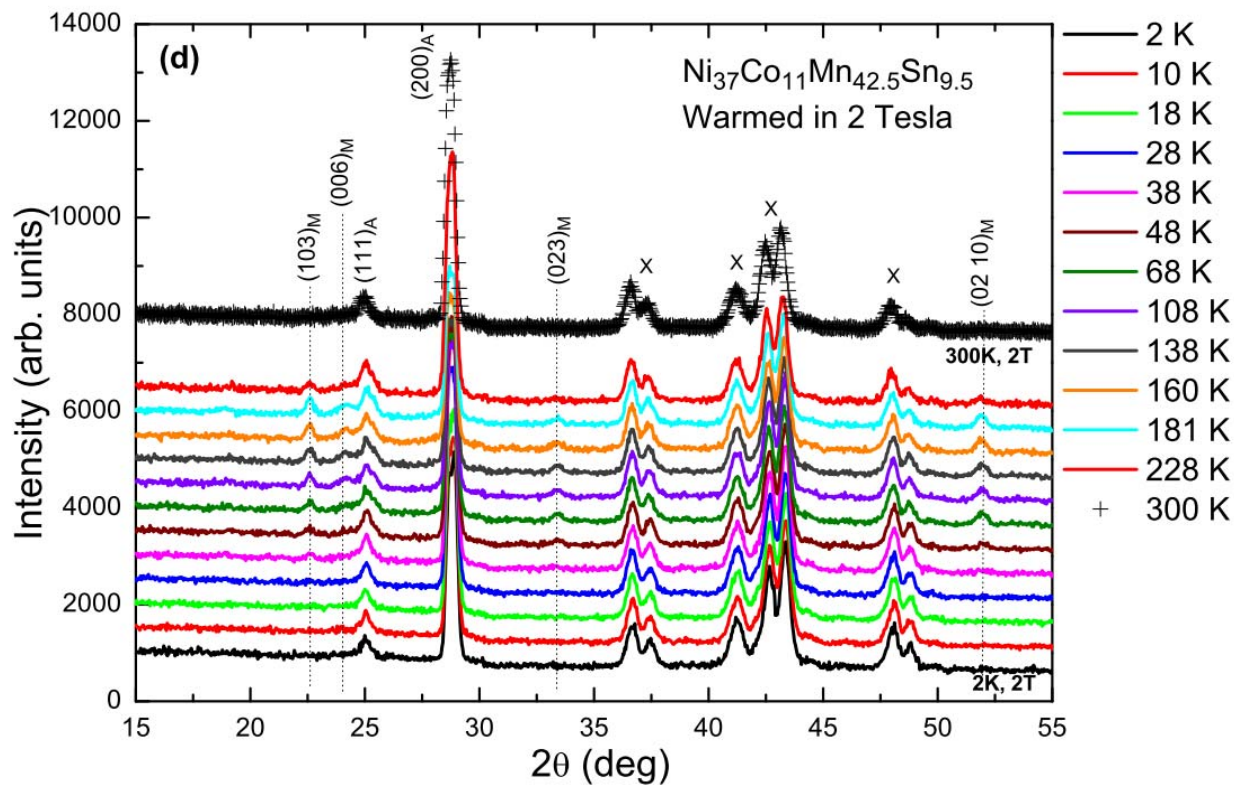

**Fig. 8(d).** Neutron diffraction patterns obtained using the CHUF protocol. The sample was cooled from 300 K to 2 K in a field of 7 Tesla which is isothermally reduced to 2 T. Data were taken in the warming cycle. Subscripts A, M and X have the same meaning as in Fig. 7.

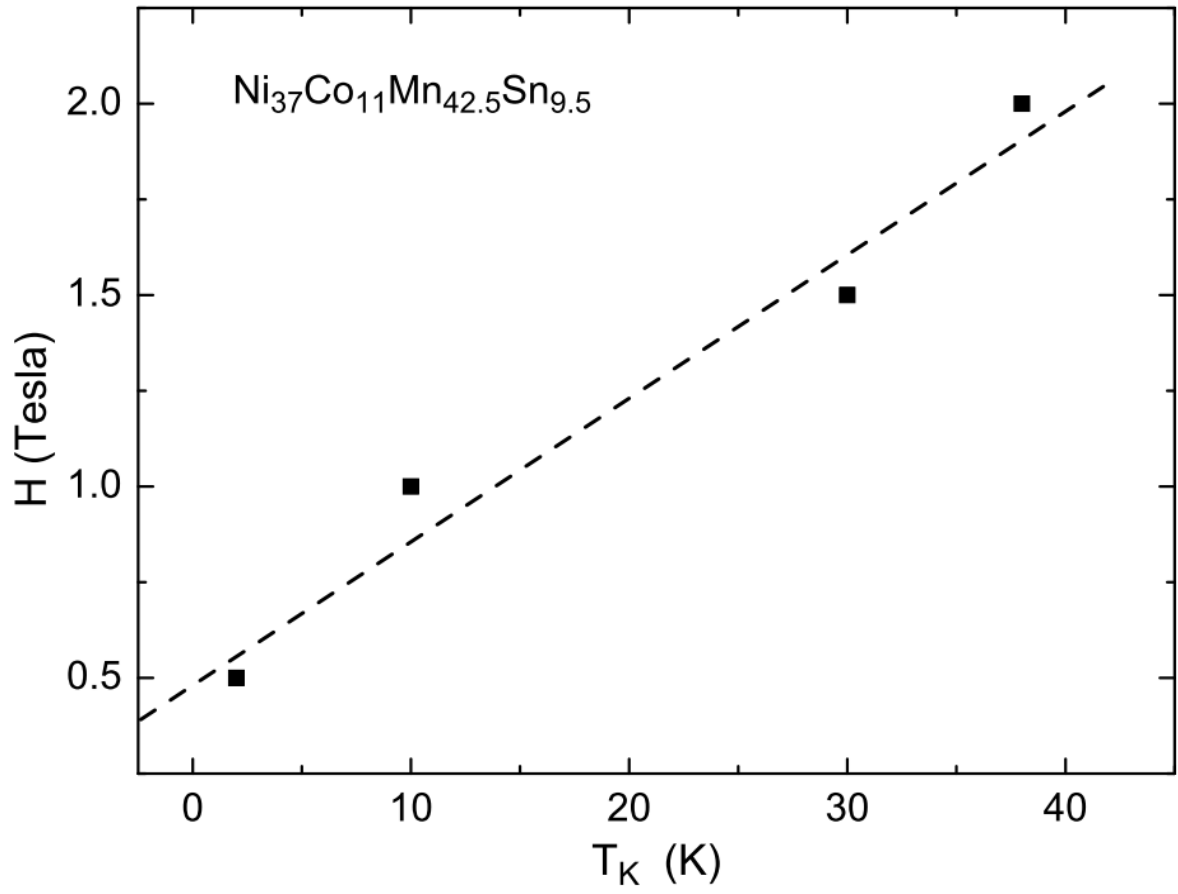

**Fig. 9.** H-T diagram showing the kinetic arrest line across which there would be a devitrification from the arrested metastable austenite phase to the equilibrium martensite phase during warming cycle.

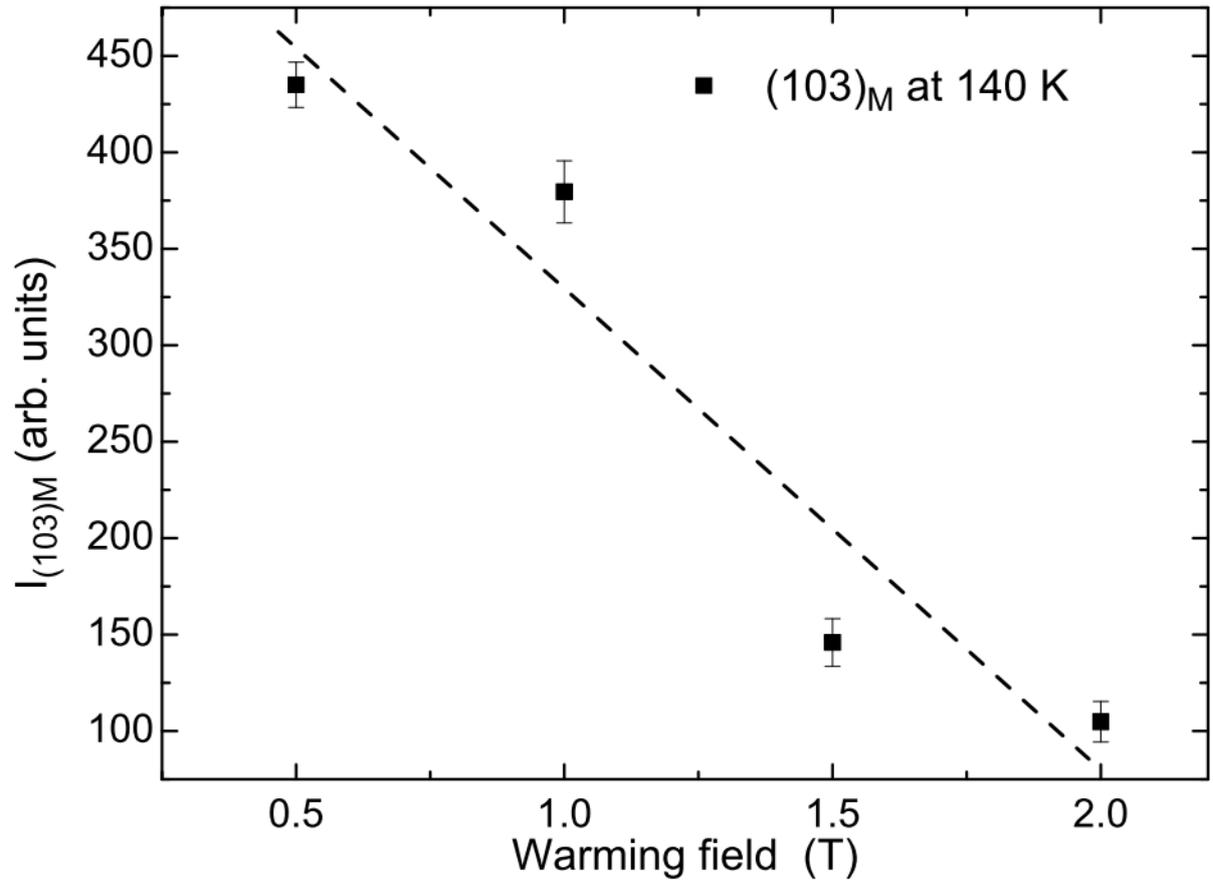

**Fig. 10.** Integrated intensity of the (103) reflection of the 10M martensite phase (obtained from Figs. 8 (a-d)) as a function of the warming field at 140 K.